\documentstyle[11pt,equation,epsfig
]{article}
\input{psfig.sty}
\begin{document}


\def\lsim{\lower2pt\hbox{$\buildrel{<}\over{\sim}$}}
\def\gsim{\lower2pt\hbox{$\buildrel{>}\over{\sim}$}}
\def\rr{{\rm r}}
\def\cc{{\rm c}}
\def\mm{{\rm m}}
\def\ss{{\rm s}}
\newcommand{\FF}{{\cal F}}
\newcommand{\cd}{\cdot}
\newcommand{\ip}{\int_0^{2\pi}}
\newcommand{\al}{\alpha}
\renewcommand{\b}{\beta}
\newcommand{\de}{\delta}
\newcommand{\De}{\Delta}
\newcommand{\ep}{\epsilon}
\newcommand{\ga}{\gamma}
\newcommand{\Ga}{\Gamma}
\newcommand{\ka}{\kappa}
\newcommand{\io}{\iota}
\newcommand{\La}{\Lambda}
\newcommand{\la}{\lambda}
\newcommand{\Om}{\Omega}
\newcommand{\om}{\omega}
\newcommand{\si}{\sigma}
\newcommand{\Si}{\Sigma}
\newcommand{\th}{\theta}
\newcommand{\vth}{\vartheta}
\newcommand{\vph}{\varphi}
\newcommand{\ra}{\rightarrow}
\newcommand{\tr}{\mbox{tr}}
\newcommand{\hor}{\mbox{hor}}
\renewcommand{\baselinestretch}{1.01}
\newcommand{\bea}{\begin{eqnarray}}
\newcommand{\eea}{\end{eqnarray}}
\newcommand{\dd}{\partial}
\def\laq{\raise 0.4ex\hbox{$<$}\kern -0.8em\lower 0.62
ex\hbox{$\sim$}}
\def\double{\baselineskip 24pt \lineskip 10pt}


\newcommand{\sx}{\sigma}
\newcommand{\sei}{\sigma_8}
\newcommand{\sxa}{\sigma_1}
\newcommand{\sxb}{\sigma_2}
\newcommand{\sxc}{\sigma_3}
\newcommand{\pha}{\phi_1}
\newcommand{\phb}{\phi_2}
\newcommand{\phc}{\phi_3}
\newcommand{\Pha}{\Phi_1}
\newcommand{\Phb}{\Phi_2}
\newcommand{\Phc}{\Phi_3}
\newcommand{\Phib}{\bar{\Phi}}
\newcommand{\Phab}{\bar{\Phi}_1}
\newcommand{\Phbb}{\bar{\Phi}_2}
\newcommand{\Phbc}{\bar{\Phi}_3}
\newcommand{\mpl}{m_{Pl}}
\newcommand{\Mpl}{M_{Pl}}
\newcommand{\lx}{\lambda}
\newcommand{\Lx}{\Lambda}
\newcommand{\ex}{\epsilon}
\newcommand{\be}{\begin{equation}}
\newcommand{\ee}{\end{equation}}
\newcommand{\een}{\end{subequations}}
\newcommand{\ben}{\begin{subequations}}
\newcommand{\beq}{\begin{eqalignno}}
\newcommand{\eeq}{\end{eqalignno}}
\def \lta {\mathrel{\vcenter
     {\hbox{$<$}\nointerlineskip\hbox{$\sim$}}}}
\def \gta {\mathrel{\vcenter
     {\hbox{$>$}\nointerlineskip\hbox{$\sim$}}}}
\pagestyle{empty}
\noindent
\begin{flushright}
\end{flushright} 

\vspace{2cm}
\begin{center}
{ \Large \bf
Spectrum of Cosmological Perturbations \\
\vspace{0.2cm}
from Multiple-Stage Inflation
} 
\\ \vspace{1cm}
{\large 
M. Sakellariadou$^{(a)}$ and N. Tetradis$^{(b)}$ 
}
\\
\vspace{1cm}
{\it
$^{(a)}$ D\'epartement de Physique Th\'eorique, Universit\'e de
Gen\`eve, \\
24 quai E. Ansermet,  CH-1211 Geneva, Switzerland
\\
\vspace{0.4cm}
$^{(b)}$ Scuola Normale Superiore, Piazza dei Cavalieri 7, 56100 Pisa, Italy
}
\\
\vspace{2cm}
\abstract{
Within the context of supersymmetric hybrid inflation, we study a 
multiple-stage inflationary scenario that can generate a
primordial spectrum of adiabatic density perturbations with a 
break at $k_b \simeq 0.05~h~{\rm Mpc}^{-1}$. The presence of such
a break is supported by the APM galaxy survey data.
We consider a specific model within this scenario 
and confront it with observational data. 
We reproduce the angular 
power spectrum of CMB anisotropies, and also account for the break 
in the power spectrum of galaxy clustering. 
In addition, we find a value for $\sigma_8$
in agreement with the one deduced from observations.
A characteristic property of the spectrum is 
a drop of the spectral index from 
$n \simeq 1$ to $n \sim 0.6$ at $k \sim k_b$. 
\\
\vspace{1cm}
}
\end{center}
\vspace{1cm}
\noindent

\newpage
\pagestyle{plain}
\setcounter{page}{1}

\setcounter{equation}{0}
\renewcommand{\theequation}{{\bf 1.}\arabic{equation}}

\section{Introduction}

The origin of the measured anisotropies in the 
cosmic microwave background
and the generation and evolution of large-scale structure in the
universe, such as the observed clustering in the galaxy 
distribution, represent outstanding questions in modern cosmology. 
Within the framework
of gravitational instability, there are two currently investigated
families of models to 
explain the formation of the observed structure.  Initial density 
perturbations can either be due to ``freezing in'' of quantum fluctuations
of a scalar field during an inflationary period 
\cite{stein}, or they may be seeded by topological defects, which can form 
naturally during a symmetry breaking phase transition in the early universe
\cite{kibble}.
The cosmic microwave background (CMB) anisotropies provide a link 
between theoretical predictions and observational data, which may allow us
to distinguish between inflationary models and defect scenarios by purely
linear analysis \cite{mairiruth}. 

Within the inflationary paradigm, a possible mechanism 
for the generation of the angular power spectrum of CMB anisotropies and the 
creation of the large-scale structure is based on the quantum fluctuations 
that exited the horizon during inflation \cite{freeze}.
These fluctuations are the source of the primordial spectrum of
density inhomogeneities \cite{denpert}, which has left an imprint on the CMB. 
The observed large-scale structure could have been generated by the growth 
through gravitational instability of this  primordial spectrum of 
perturbations in the otherwise uniform  distribution of matter.

By making some assumptions about the matter content of the early universe, 
one can investigate whether
a given spectrum of primordial perturbations 
can evolve into the large-scale structure observed today.
The evolution of the spectrum depends crucially on the nature and amount of the
dark matter in the universe. The two extreme cases, 
of hot dark matter (HDM)
and of cold dark matter (CDM), lead to inconsistencies 
with observational data. 
The HDM model has been abandoned, since the thermal motion of massive neutrinos
wipes out any small-scale structure \cite{wfd83}. On the other hand, 
the standard
CDM model with a flat (spectral index $n=1$) initial spectrum of adiabatic
perturbations predicts too much power on small scales, 
if one normalizes it to the
COBE data at large scales. The standard CDM model leads to a value 
$\sigma_8 \simeq 1.2$ for the variance
of the total mass fluctuation in a sphere of radius $8~{\rm Mpc}/h$
\cite{dp83}. 
On the other hand, the observational value, inferred from
the abundances of rich clusters of galaxies,
is $\sigma_8\simeq 0.6\pm 0.2$ \cite{wef93,viana}. 

As the standard CDM model is a generally successful model 
(e.g., it successfully
reproduces galaxy clustering statistics, the various epochs 
of structure formation,
as well as peculiar velocity flows, while it maintains 
CMB anisotropies at a level
lower than observational upper limits), several approaches have been proposed,
which modify its underlying assumptions (for a review see Ref.~\cite{report}).
In order to render CDM models compatible with observational data, 
one possibility is to change the
initial spectrum of perturbations by introducing a tilted (spectral
index $n<1$) scale-free spectrum. However, this 
approach has not led to successful predictions
\cite{tilt}.
Another approach is to 
consider less standard values for the cosmological parameters, 
while leaving the
initial spectral index unchanged (i.e., $n\simeq 1$). 
For instance, one can either
investigate CDM models with a higher fraction of baryons \cite{bar}, 
or models which have a mixture of CDM with an amount
of HDM, or even  models with a positive cosmological constant 
\cite{cosmconst,lambda}.

Another interesting possibility is to 
consider power spectra with a break,  
which arise naturally
in models with more than one stages of inflation -- an approach 
adopted by several authors \cite{doubleinf}--\cite{subir}. 
A model of double chaotic
inflation which generates such a spectrum has been studied in
detail in Refs.~\cite{pol1,pol2,pol3}. 
The model contains two scalar fields 
with disconnected quadratic potentials.
 
In this paper, we discuss the possibility of a multiple-stage
inflationary scenario within the context of supersymmetric 
hybrid inflation \cite{cop}. The necessary superpotential 
can arise in supersymmetric GUT models. Moreover,
the observable part of inflation takes place for field values
below the Planck scale, so that supergravity corrections 
are under control. Inflationary 
models such as the one we are considering here have
been discussed in Ref.~\cite{mecostas}, and have been shown
to survive the supergravity corrections in Ref.~\cite{megeorge}. 

A strong motivation for these models has its origin in the
issue of the initial conditions \cite{first1}--\cite{nikos} 
that are necessary 
for hybrid inflation \cite{hybrid}
\footnote{An alternative mechanism for the resolution of this problem
has been proposed in Ref.~\cite{mezurab}.
}. 
This is a consequence of the 
presence of (one or more) scalar fields orthogonal to the inflaton,  
and the constraint from the COBE data on 
the inflationary energy scale
$V^{1/4}$ 
(determined by the vacuum energy density during inflation), which 
must be at least two or three orders of magnitude smaller
than the Planck scale \cite{lyth}
\footnote{Throughout the paper we use the ``reduced'' Planck
scale 
$\mpl= \Mpl/\sqrt{8 \pi}$, $\Mpl = 1.22 \times 10^{19}$ GeV.}.
The difference between the energy scale $\mpl$ at the end of 
the Planck era, when classical general relativity becomes applicable,
and the inflationary scale $V^{1/4}$ implies that 
the various fields evolve for a long time before settling down along the 
inflationary trajectory. The Hubble parameter $H$ 
sets the scale for the ``friction'' term in the field evolution equations. 
When the energy density drops much below $\mpl^4$, the
smallness of this ``friction'' term results in a very long  
evolution, during which the fields oscillate around zero 
many times. Some of the trajectories eventually 
settle down in the valley of the potential that produces inflation.
However, the sensitivity to the initial conditions is 
high because of the long evolution. A slight variation of
the initial field values separates inflationary trajectories from 
trajectories that lead to the 
minima of the potential, where inflation does not occur.
As a result, the initial field configuration must be extremely fine-tuned
for inflation to set in. The fields orthogonal to the inflaton
(whose typical fluctuations at the end of the Planck era are $\sim \mpl$)
must be zero with an accuracy of at least $\sim 10^{-5}\mpl$
over regions that exceed the initial Hubble length by one or two
orders of magnitude \cite{nikos}.

In Ref.~\cite{mecostas}
a simple resolution of the issue of fine-tuning
described above was suggested.
A scenario
with two stages of inflation was proposed within the context
of global supersymmetry. 
The first stage has a typical scale
$\sim \mpl$, which implies that the ``friction'' term
in the field evolution equations of the fields is large.
As a result, the system settles down quickly along an almost 
flat direction  of the potential 
and this stage of inflation
occurs naturally. By generating an exponential expansion of the 
initial region of space, 
it also provides the homogeneity that 
is necessary for the second stage. The latter has a characteristic
scale much below $\mpl$ and generates the 
density perturbations 
that result in the CMB 
anisotropy observed by COBE. This scenario was 
generalized in the context of supergravity in Ref.~\cite{megeorge}.

It is a logical next step to contemplate the 
possibility of a multiple-stage inflationary scenario,
with a sequence of scales that 
starts near $\mpl$ and continues down to the scale
implied by COBE, or even to lower energy scales.
For inflation driven by one field in the context of supergravity,
such a possibility
has already been proposed in Ref.~\cite{subir} and has been
associated with a break 
at $k_b \simeq 0.05~h~{\rm Mpc}^{-1}$
in the power spectrum 
of galaxy clustering derived from the APM galaxy survey.
This break has been linked to a possible phase
transition between two of the inflationary stages.
In this paper we would like to provide another 
realization of such a scenario, within a multi-field theory, with an estimation
of the predicted power spectrum. The determination of the spectrum during
the phase transition is beyond our technical capabilities at the
moment. However, the phase transition affects only a small 
range of scales, while the rest of the power spectrum is strongly
constrained by the observational data.

Firstly, we are interested in examining whether the COBE data 
can be made compatible with the data of galaxy surveys, 
employing the smallest number of cosmological parameters.  
More specifically, we consider
a CDM model, with zero cosmological constant, present Hubble parameter 
$h=0.5~~(H_0 \equiv h~100~{\rm km/s/Mpc})$ 
and flat geometry $\Omega_{matter}=1$. 
We assume that the fraction of the critical density in baryons is
$\Omega_b=0.05$ and in cold dark matter $\Omega_{CDM}=0.95$.
Subsequently, we repeat the calculation, 
allowing for a non-zero value of the cosmological constant, a scenario 
supported by recent observations \cite{lambda}.

In Section 2 we discuss the general form of the power spectrum
predicted by the APM data. In Section 3 we present our model and
in Section 4 the properties of the inflationary stages. 
The primordial power spectrum in our model is calculated in Section 5
and is compared with observations in Section 6. Our conclusions
are given in Section 7.

\setcounter{equation}{0}
\renewcommand{\theequation}{{\bf 2.}\arabic{equation}}

\section{The APM power spectrum}

An indication for the possibility of more than one stages of
inflation is provided by the power spectrum deduced from the 
APM galaxy survey \cite{apm}. This spectrum has a break at 
a characteristic scale $k_b \simeq 0.05~h~{\rm Mpc}^{-1}$
\cite{data1,peac,subir,brdat2}. One explanation for this break 
could be a sharp change of the spectral index of
the primordial spectrum between two stages of inflation \cite{subir}. 
In this section we summarize the main points of the analysis in the above
references, in order to motivate the multiple-stage scenario. 
In the following sections we shall show how such a scenario can be realized in
the context of a specific model.

In Fig.~1 we plot the logarithmic slope of the power spectrum 
recovered from the APM angular galaxy catalogue \cite{data1,brdat2}.
(We have used the data of Table~2 in Ref.~\cite{brdat2}.) 
The slope has been calculated through a least-squares fit for groups of 3
data points, assuming that the spectrum is locally linear in a logarithmic 
scale. The results are in agreement with
the detailed analysis of Ref.~\cite{brdat2}, even though our estimated errors
are slightly different. A sharp change in the slope of the spectrum
is visible around $k_b \simeq 0.05~h~{\rm Mpc}^{-1}$.

This break persists even after the effects of the non-linear evolution
of matter fluctuations (when the density contrast becomes of order 1) 
have been removed. Through $N$-body simulations,
Baugh and Gazta\~naga \cite{brdat2} have estimated the 
linear spectrum that gives rise to the observed spectrum for
a spatially flat universe with critical matter density and 
zero cosmological constant. Their estimate
is well fitted by the curve \cite{brdat2,linear}
\be
P_l(k) = { C k^\alpha \over [ 1 + (k/k_b)^2 ] ^\beta},
\label{lfit} \ee
with $C \simeq 7.0\times 10^5, ~~k_b \simeq 0.05~h~{\rm Mpc}^{-1},
~~\alpha \simeq 1, ~~\beta \simeq 1.6$. This result is in agreement with the
spectrum obtained through the application of the formula of
Jain {\it et al.} \cite{jmw} for the relation between the linear and
non-linear spectra. The 
formula of Peacock and Dodds \cite{pd} does not lead to 
very good agreement with the results of the
$N$-body simulations \cite{linear}. In Fig.~1 we display the 
logarithmic slope of the linear spectrum as predicted by 
Eq.~(\ref{lfit}) (solid line). The sharp decrease in the slope 
is again apparent.

The implications for the primordial spectrum can be obtained if one
deconvolutes the spectrum by using the linear transfer function, 
which determines the scale-dependent growth of linear perturbations. 
A parametrization of this function is given by the relation \cite{ge}
\be
T(k) = \left[ 1 + \left\{
a k + (bk)^{3/2} + (ck)^2
\right\}^{\nu}
\right]^{-1/\nu},
\label{tfit} \ee
with $a = 6.4~\Gamma^{-1}h^{-1}{\rm Mpc}$,
$b = 3~\Gamma^{-1}h^{-1}{\rm Mpc}$,
$c = 1.7~\Gamma^{-1}h^{-1}{\rm Mpc}$,
$\nu=1.13$ and $\Gamma = \Omega h e^{-2\Omega_b}$, where
$\Omega_b$ is the fraction of critical density in baryonic matter. 
For $h=0.5$, $\Omega_b=0.05$, the spectral index $n$ of the 
primordial spectrum $P(k) \sim P_l(k)/T^2(k)$
is given by the dashed line in Fig.~1. A transition from a value
$n\simeq 1$ at large scales to an average 
value $n \sim 0.5$ at small scales is observed. It must be pointed out
that the above analysis cannot be considered conclusive, as the parametrization
of Eq. (\ref{tfit}) may not be sufficiently accurate. In section 6 we 
present a detailed comparison with the linear power spectrum of 
Eq. (\ref{lfit}) using the code CMBFAST of Seljak and Zaldarriaga
\cite{cmbfast}.

The variation of the spectral index can be explained by the generation of 
density perturbations during 
two stages of inflation, driven by fields with different potentials.
The first stage must have lasted long enough for scales between
$k \sim 3 \times 10^{-4}~h~{\rm Mpc}^{-1}$ (relevant for the 
COBE data) and $k \sim k_b \simeq 5 \times 10^{-2}~h~{\rm Mpc}^{-1}$
to have crossed outside the horizon during its duration.
This requires at least 5 e-foldings to be generated by the first stage
of inflation. 
The large deviation of the spectral index from 1 
has strong implications for the duration of the second stage. For the
models of interest to us, the inflationary potential 
has the general form $V \simeq V_0(1 + c f(\phi/\mpl))$, where $c$ is a small
constant. Moreover, the inflaton field is constrained to 
be $\phi \lta \mpl$. A general analysis of the spectral index predicted 
by such potentials, for various choices of the function $f(\phi/\mpl)$,
is given in Ref.~\cite{lyth}. According to Table~1 of that reference,
the total number of e-foldings $N$ is constrained by the relation
$|n-1|(N/50) \lta 0.08$. For $n \simeq 0.6$ we infer 
$N \lta 8$. More specifically, for the case $f(\phi/\mpl)=\ln(\phi/\mpl)$
that is relevant for our model, 
$|n-1|(N/50) \simeq 0.02$ and $N \simeq 2$.

The above considerations support a picture of multiple
short bursts of inflation, with significant variations of 
the spectral index of
the primordial spectrum. The observational
data provide information on two of these inflationary stages.
The COBE-DMR measurements require a stage with $n\simeq 1$ and at least 
5 e-foldings, while the APM galaxy survey data support the possibility of
a second stage with $n \sim 0.6$ that generates $\sim 2$ e-foldings.
The total number of e-foldings during all the inflationary 
stages with observable consequences must be $\sim 60$ for the flatness and 
horizon problems of
standard cosmology to be resolved. However, the stages 
responsible for the last $\sim 50$ e-foldings generate density perturbations
at scales that 
are strongly affected by the non-linear evolution. As a result, it is very
difficult to extract information on their characteristics.

The above picture contrasts with the double inflationary scenario
of Refs.~\cite{pol1,pol2,pol3}, where inflation is
driven by two disconnected quadratic potentials for the fields 
$\sxa$, $\sxb$. The first stage of inflation occurs
for $\sx_{1,2} \gg \mpl$ and terminates when $\sxa \sim \mpl$, while
the second one terminates when $\sxb \sim \mpl$. For the second
stage to generate less than $\sim 60$ e-foldings, so that the break
in the spectrum is observable, the value of
$\sxb$ at its beginning must not be very far from $\mpl$. 
In our model there is no need for a similar adjustment.
The requirement of a spectral index significantly below 1 
for the part of the
spectrum generated by the second stage of inflation automatically constrains
the number of e-foldings from this stage to be less than 60, independently
of the value of $\sxb$.

In the following section we describe a model which provides the
potential for the two stages of inflation that are relevant for
the COBE and APM data. 

\setcounter{equation}{0}
\renewcommand{\theequation}{{\bf 3.}\arabic{equation}}

\section{The model}
The model we consider is described by the
superpotential 
\be
W = S_1 \left( -\mu_1^2  
+ \lx_1 \Phab \Pha  + g \Phbb \Phb \right)
+S_2 \left( -\mu_2^2 + \lx_2 \Phbb \Phb \right).
\label{two1} \ee
The superfields $\Pha$, $\Phab$ and $\Phb$, $\Phbb$
transform under internal gauge symmetry groups ${\rm \cal M}_1$
and ${\rm \cal M}_2$ respectively.
The  $S_1$, $S_2$ superfields are gauge singlets. 
We have assumed that $S_2$ is the linear combination of gauge
singlets that does not couple to 
$\Phab \Pha$ \cite{mecostas}.
We make the simplifying assumption 
that the scalar components of the various superfields
are real. The imaginary components do not alter the qualitative
picture of inflation in this model, while they would make the 
calculation of the spectrum of density perturbations 
much more complicated.
Staying 
along the $D$-flat directions, we define canonically normalized
scalar fields according to 
\footnote{
We use the same notation for the superfields and their scalar 
components.
}
\beq
S_1 =&~\frac{\sxa}{\sqrt{2}},
~~~~~~~~~~~~~~~~~~~
\Pha = \Phab = ~\frac{\pha}{2},
\nonumber \\
S_2 =&~\frac{\sxb}{\sqrt{2}},
~~~~~~~~~~~~~~~~~~~
\Phb = \Phbb = ~\frac{\phb}{2}.
\label{two2} \eeq

The potential is then given by the expression 
\beq
V(\sxa,\sxb,\pha,\phb)=\sum_i
\left|\frac{\partial W}{\partial \Phi_i} \right|^2
=&~
\left( \mu_1^2 - \frac{\lx_1}{4} \pha^2 \right)^2
-\frac{g}{2}\mu_1^2\phb^2 +\frac{g^2}{16}\phb^4 
+\frac{g\lx_1}{8} \pha^2 \phb^2 
+ \frac{\lx_1^2}{4}\sxa^2 \pha^2 
\nonumber \\
&
+ \left( \mu_2^2 - \frac{\lx_2}{4} \phb^2 \right)^2
+ \frac{1}{4}\left(g\sxa+\lx_2\sxb\right)^2 \phb^2,
\label{two3} \eeq
where the mass scales $\mu_1$, $\mu_2$ are chosen 
$\mu_1 > \mu_2$.
The minima of this potential are located at
\be
\sxa=\sxb=0,~~~~~~~~~~~~~~
\pha^2= \frac{4}{\lx_1}\mu_1^2-\frac{4g}{\lx_1\lx_2}\mu_2^2,~~~~~~~~~~~~~~
\phb^2= \frac{4}{\lx_2}\mu_2^2.
\label{two4} \ee
For $\pha=\phb=0$ the potential is independent of 
$\sx_{1,2}$. Its value
$V=\mu_1^4+\mu_2^4$ gives the vacuum energy density during
the first stage of inflation. 
Along this direction, 
the mass terms of the $\phi_{1,2}$ fields are
\beq
M^2_{\pha}=&~
-\lx_{1} \mu_{1}^2 + \frac{\lx_{1}^2}{2} \sxa^2,
\label{two5a} \\
M^2_{\phb}=&~
-\lx_{2} \mu_{2}^2 - g \mu_1^2
+ \frac{1}{2} \left(g \sxa +\lx_2 \sxb \right)^2.
\label{two5b} \eeq
The mass term $M^2_{\pha}$ becomes negative for 
\be
\sxa^2  < \sx^2_{1ins} = \frac{2 \mu_1^2}{\lx_1}.
\label{two6} \ee
This indicates the presence of an instability which can lead to the
growth of the $\phi_{1}$ field.

Another flat direction (independent of $\sxb$)
exists for $\sxa=0$, $\pha^2= 4\mu_1^2/\lx_1$, $\phb=0$.
The value of the potential
$V=\mu_2^4$ gives the vacuum energy density during
the second stage of inflation. 
The mass term of the $\phb$ field is given by 
\be
M^2_{\phb}=~
-\lx_{2} \mu_{2}^2 + \frac{\lx_{2}^2}{2} \sxb^2.
\label{two6a} \ee
An instability appears for
\be
\sxb^2  < \sx^2_{2ins} = \frac{2 \mu_2^2}{\lx_2},
\label{two6b} \ee
which can lead to the
growth of the $\phi_2$ field.

The flatness of the potential is lifted by radiative corrections.
During the first stage of inflation and 
for $\sx_{1,2}$ far above the instability points,
the one-loop 
contribution to the effective potential is
\footnote{
It must be pointed out that the supersymmetric cancellations 
that lead to Eqs.~(\ref{two7}), (\ref{two8})
require the inclusion
of radiative corrections from both the real and imaginary scalar 
components of
the superfields $\Pha$, $\Phb$. However, 
this is the only effect of the
imaginary components on 
the inflationary stages. This is the reason why we omitted them
in Eq.~(\ref{two3}). For a discussion that includes these
components see Ref.~\cite{mecostas}.
}
\beq
\Delta V(\sxa,\sxb) \simeq&~ 
\frac{M_1}{16 \pi^2}  \lx_1^2 \mu_1^4
\left[ \ln\left(
\frac{\lx_1^2 \sxa^2}{2 \Lambda_1^2} 
\right) 
+ \frac{3}{2}
\right]
\nonumber \\
&+ \frac{M_2 }{16 \pi^2}  
\left( \lx_2 \mu_2^2 + g \mu_1^2\right)^2
\left[ \ln\left(
\frac{\left(g\sxa + \lx_2 \sxb\right)^2}{2 \Lambda_2^2} 
\right) 
+ \frac{3}{2}
\right],
\label{two7} \eeq
where $M_{1,2}$ are the dimensionalities of the representations
of the groups ${\rm \cal M}_{1,2}$ to which the superfields
$\Phi_{1,2}$ belong. The exact value of the normalization scales
$\Lx_{1,2}$ is not important for our discussion.
In the following we shall consider theories in which 
the two sectors $(S_1,\Pha)$ and $(S_2,\Phb)$ are essentially decoupled. 
For this reason we shall assume that the coupling $g$ is sufficiently 
small for the radiative correction of Eq.~(\ref{two7}) to be approximated 
as
\be
\Delta V(\sxa,\sxb) \simeq~ 
\frac{M_1}{16 \pi^2}  \lx_1^2 \mu_1^4
\left[ \ln\left(
\frac{\lx_1^2 \sxa^2}{2 \Lambda_1^2} 
\right) 
+ \frac{3}{2}
\right]
+ \frac{M_2 }{16 \pi^2}  
\left( \lx_2 \mu_2^2 + g \mu_1^2\right)^2
\left[ \ln\left(
\frac{\lx^2_2 \sxb^2}{2 \Lambda_2^2} 
\right) 
+ \frac{3}{2}
\right].
\label{two7a} \ee
The above contribution provides the slope that leads to the slow rolling
of the $\sx_{1,2}$ fields during the first stage of inflation.

During the second stage of 
inflation the slope for the $\sxb$ field is provided by the 
radiative correction
\be
\Delta V(\sxa,\sxb) \simeq 
\frac{M_2}{16 \pi^2}   \lx_2^2 \mu_2^4
\left[ \ln\left(
\frac{\lx_2^2 \sxb^2}{2 \Lambda_1^2} 
\right) 
+ \frac{3}{2}
\right].
\label{two8} \ee
Away from the flat directions the radiative corrections are
small and we neglect them.

An important question concerns the supergravity corrections to the
above potential. In particular, it is important that the flat directions
of the potential are not lifted by these corrections \cite{cop}.
A detailed analysis of this problem is presented in Ref.~\cite{megeorge},
where it is shown that, for the flat directions to be preserved, 
at least one of the two inflationary stages must be driven by
$D$-term energy density \cite{dterm}. However, the form of the potential
is the same as the one we described above. Two flat directions
exist with a small slope generated by logarithmic radiative corrections.

\setcounter{equation}{0}
\renewcommand{\theequation}{{\bf 4.}\arabic{equation}}

\section{The inflationary stages}

The model we described in the previous section allows for
two stages of inflation separated by an intermediate stage.

\subsection{The first stage of inflation}
The Hubble parameter during the first stage of inflation
is almost constant
\be
H^2_1 \simeq \frac{\mu_1^4+\mu^4_2}{3 \mpl^2}.
\label{three1} \ee
We assume that the part of this stage with
observable consequences starts at a time $t_{1i}$ and
finishes at a time $t_{1f}$.
Because of our assumption $\mu_1 > \mu_2$, the vacuum
energy density associated with the fields $\sxa$, $\pha$
dominates the first stage. 
Therefore, this stage terminates when the ``slow-roll'' conditions 
for $\sxa$ are not satisfied any more. 
For the potential of Eqs.~(\ref{two3}), (\ref{two7a}), this happens for 
\be
\sxa^2\left(t_{1f}\right) \equiv \sx_{1f}^2 \simeq 
\frac{M_1 \lx^2_1}{8 \pi^2} \mpl^2. 
\label{three2} \ee
Unless the coupling $\lx_1$ is taken much smaller than 1,
$\sx^2_{1f}$ is much larger than the instability point 
$\sx^2_{1ins}$ of Eq.~(\ref{two6}).

The number of e-foldings from a time $t$ until the end of the first stage is
given by 
\be
N_1(t)=\ln\left(\frac{a_{1f}}{a(t)} \right)=
\frac{4\pi^2}{M_1\lx_1^2} \frac{\mu_1^4+\mu^4_2}{\mu^4_1}  
\frac{\sxa^2(t)-\sx_{1f}^2}{\mpl^2},
\label{three3} \ee
where $a(t)$ is the scale factor and $a_{1f}\equiv a\left( t_{1f} \right).$
The total number of e-foldings during the observable part of 
this stage is given by
\be
N_{1tot}=\ln\left(\frac{a_{1f}}{a_{1i}} \right)
=\frac{4\pi^2}{M_1\lx_1^2} \frac{\mu_1^4+\mu^4_2}{\mu^4_1}  
\frac{\sx_{1i}^2-\sx_{1f}^2}{\mpl^2},
\label{three4} \ee
with 
$\sx_{1i}^2 \equiv \sxa^2\left(t_{1i}\right)$,
$a_{1i} \equiv a\left(t_{1i}\right)$.
The relative change of the values of the two fields during the 
first stage is 
\be
\frac{\sxb^2\left(t_{1i} \right) - \sxb^2\left(t_{1f} 
\right)}{\sxa^2\left(t_{1i} \right) - \sxa^2\left(t_{1f} 
\right)} =
\frac{\sxb^2\left(t_{1i} \right) - \sxb^2\left(t_{1f} 
\right)}{\sx^2_{1i}  - \sx^2_{1f} }
= 
\frac{M_2 }{M_1} 
\frac{\left(\lx_2 \mu^2_2+g\mu^2_1\right)^2}{ \lx_1^2 \mu^4_1}.
\label{three5} \ee
In the following, we shall use parameters
that make this ratio much smaller than 1, so that the evolution
of $\sxb$ can be neglected during the first stage of inflation.

\subsection{The intermediate stage}
The intermediate stage lasts between times $t_{1f}$
and $t_{2i}$. After the end of the slow-roll regime,
the $\sxa$ field quickly rolls beyond the instability point
of Eq.~(\ref{two6}).
Subsequently, 
large domains start appearing in which the value of the 
$\pha$ field grows exponentially with time. 
For statistical systems, for which the expansion of the 
universe is not relevant, 
this process is characterized as spinodal decomposition.
The expansion of the universe complicates the above picture,
but the details are not important for our discussion. 
We assume that this initial stage of instability is fast, and soon 
the fields 
take values away from the $\sxa$ axis and 
in the vicinity of  
the minimum at $\sxa=0$, $\pha^2 = 4\mu^2_1/\lx_1$, where
the curvature of the potential is positive.
Our assumption is reasonable because the $\sxa$ field 
rolls to the origin within a time $\sim H_1^{-1} \simeq 
\sqrt{3} \mpl / \sqrt{ \mu^4_1+\mu^4_2}$ or slightly larger.
On the other hand, the 
typical time scale for the growth of the 
$\pha$ field is given by the absolute value of the 
curvature at the origin and is $\sim \left( \sqrt{\lx_1} \mu_1\right)^{-1}$. 
As a result, we expect that 
$\pha$ grows to a value near the minimum within 
a fraction of a Hubble time. 

After a short complicated evolution,
the massive fields $\sxa$, $\pha$ 
settle into
a regular oscillatory pattern around the minimum, 
with the universe 
characterized by an equation of
state $p=w \rho$. 
For a system of massive oscillating fields, such as the one
we are considering, 
$w = 0$.
From this point on,
the energy density of the oscillating fields 
is dissipated through the expansion of the
universe.
If the fields have decay channels into lighter 
species that eventually thermalize,
the equation of state of the 
radiation-dominated universe has 
$w = 1/3$.
When the energy density becomes comparable to 
$\mu_2^4$ the second stage of inflation can begin.
During the intermediate stage, the scale factor increases by
a total amount 
\be
N_{int}=\ln\left(\frac{a_{2i}}{a_{1f}} \right)
= \frac{2}{3(1+w)} \ln \left(
\frac{H_1}{H_2} \right),
\label{la} \ee
where $a_{2i} \equiv a\left(t_{2i}\right)$.

The effect of the oscillations of $\sxa$, $\pha$ on the 
stability of $\sxb$, $\phb$ is minimized if 
the former have fast decay channels into lighter species. 
Under this assumption, the fields $\sxb$, $\phb$ remain
constant during the intermediate stage, even when the 
supergravity corrections are taken into account \cite{megeorge}.
In the following we make this assumption and use $w=1/3$ during
the whole intermediate stage.

\subsection{The second stage of inflation}
This stage starts at a time $t_{2i}$ and
finishes at a time $t_{2f}$.
The Hubble parameter is
\be
H^2_2 \simeq \frac{\mu^4_2}{3 \mpl^2}.
\label{three6} \ee
Inflation stops when the ``slow-roll'' conditions 
for $\sxb$ are not satisfied any more. 
For the potential of Eqs.~(\ref{two3}), (\ref{two8}), this happens for 
\be
\sxb^2\left(t_{2f}\right) \equiv \sx_{2f}^2 \simeq 
\frac{M_2 \lx^2_2}{8 \pi^2} \mpl^2. 
\label{three7} \ee
Unless the coupling $\lx_2$ is taken much smaller than 1,
$\sx^2_{2f}$ is much larger than the instability point 
$\sx^2_{2ins}$ of Eq.~(\ref{two6}).

The number of e-foldings from a time $t$ until the end of the second stage is
given by 
\be
N_2(t)=\ln\left(\frac{a_{2f}}{a(t)} \right)
=\frac{4\pi^2}{M_2\lx_2^2} 
\frac{\sxb^2(t)-\sx_{2f}^2}{\mpl^2},
\label{three8} \ee
where $a_{2f}\equiv a\left( t_{2f} \right).$
The total number of e-foldings during this stage is given by
\be
N_{2tot}=\ln\left(\frac{a_{2f}}{a_{2i}} \right)=
\frac{4\pi^2}{M_2\lx_2^2} 
\frac{\sx_{2i}^2-\sx_{2f}^2}{\mpl^2},
\label{three9} \ee
with 
$\sx_{2i}^2 \equiv \sxb^2\left(t_{2i}\right)$.

In the next sections we shall consider parameters for which
the above equation predicts $N_{2tot}\simeq 2$ e-foldings.
As the accuracy of the analytical expressions presented in
this section is questionable for so small $N_{2tot}$,
we have verified the properties of the second inflationary 
stage numerically. 
We have found a value for $N_{2tot}$ that is close to 3 for our
choice of parameters. 

\subsection{The remaining stages of inflation}

The total number of e-foldings during all the inflationary 
stages with observable consequences must be $\sim 60$ for the flatness and 
horizon problems of
standard cosmology to be resolved. This means that additional inflationary
stages must
take place after the first two that we discussed above.
 
A simple extension of our model that would provide this possibility
involves an additional term  $S_3 
\left( -\mu_3^2 + \lx_3 \bar{\Phi}_3 \Phi_3 \right)$ in the superpotential.
We assume that the new system of superfields $S_3$, $\bar{\Phi}_3$, $\Phi_3$
is decoupled from the ones driving the first two stages of inflation.
The potential and the properties of the third inflationary
stage are completely analogous to the ones we described above.
For a sufficiently small scale $\mu_3$ the discussion of the
previous subsections is not affected. A third stage of inflation starts after
a second intermediate stage, as soon as the energy density becomes
comparable to $\mu^4_3$. The total number of e-foldings 
during the third stage is 
\be
N_{3tot}=\ln\left(\frac{a_{3f}}{a_{3i}} \right)=
\frac{4\pi^2}{M_3\lx_3^2} 
\frac{\sx_{3i}^2-\sx_{3f}^2}{\mpl^2},
\label{three99} \ee
where
$\sx_{3f}^2/\mpl^2 \simeq M_3 \lx^2_3/8\pi^2$ and the various quantities are
defined in analogy to the first two stages.
One can envisage several stages driven by such systems of superfields.
The calculation of the supergravity corrections to the potential of such
complicated systems of fields is a difficult task. However, we assume that
it is possible to guarantee the presence of the necessary number of
flat directions, even when supergravity corrrections are taken into account.

There are other possible sources of inflation as well. The increase of
temperature during the reheating after the first inflationary stages may
lead to the restoration of spontaneously broken symmetries. Additional
stages of inflation may take place when the temperature falls below the
energy scale associated with these symmetries. Inflationary stages, 
such as the thermal inflation
of Ref. \cite{thermal}, may occur even as low as at the TeV scale.

\setcounter{equation}{0}
\renewcommand{\theequation}{{\bf 5.}\arabic{equation}}

\section{The primordial spectrum}

The primordial spectrum of
density inhomogeneities has its origin in the 
quantum fluctuations that crossed outside the horizon   
during inflation \cite{freeze}.
The scale $k$ crosses the Hubble radius when
$k=aH$.
The scale $k_{COBE}/a_0 \sim H_0$ 
exited the horizon at the beginning of the observable part of the 
first stage of inflation. 
This implies that
$a_0H_0 = a_{1i} H_1$. 
(Here $a_0$, $H_0\simeq \left(3000~{\rm Mpc}\right)^{-1}h $ 
are the present values of 
the scale factor and the Hubble parameter. Following the standard
convention, we set $a_0=1$ in the following.)
 
For $k\lta k_2=a_{2i}H_2$ the Hubble radius is 
crossed only once, during the first stage of inflation, while 
for $k \gta k_1=a_{1f}H_1$ it is 
crossed once during the second stage.
The scales $k_2<k < k_1$, however, cross the Hubble radius
three times: They exit the horizon during the first stage, re-enter
duing the intermediate stage and exit again during the second stage.
We define the scale $K$ according to 
\be
K= \sqrt{k_1 k_2}
\label{four2} \ee
and we obtain 
\be
\ln \left( \frac{K}{H_0} \right)=N_{1tot}+\frac{1}{2}N_{int}
+\frac{1}{2} \ln \left( \frac{H_2}{H_1} \right),
\label{four3} \ee
with 
$N_{1tot}$, $N_{int}$ given by Eqs.~(\ref{three4}), (\ref{la}) respectively.
For the ratio $k_2/k_1$ we obtain
\be
\ln \left( \frac{k_2}{k_1} \right) = \left( 1 - \frac{2}{3(1+w)}\right)
\ln \left( \frac{H_2}{H_1} \right).
\label{four4} \ee

The gravitational potential for scalar metric perturbations that are generated
during the first stage of inflation is given by the expression
\cite{pol1,doublen,mukhanov}
\be
\Phi =~2C \frac{\dot{H}}{H^2} + D \frac{1}{H}
\frac{V_1\dot{V}_2-\dot{V}_1V_2}{2\mpl^2\left(V_1+V_2\right)}, 
\label{four5} \ee \\
with
\beq
C=&~-\frac{1}{2}\left[ H
\left( \frac{V_1}{V_1+V_2} \frac{\delta\sxa}{\dot{\sx}_1}
+\frac{V_2}{V_1+V_2} \frac{\delta\sxb}{\dot{\sx}_2}
\right) \right]_{k=aH}
\label{four6} \\
D=&~\left[ \frac{1}{3H}\left( 
\frac{\delta\sxa}{\dot{\sx}_1}-\frac{\delta\sxb}{\dot{\sx}_2}
\right) \right]_{k=aH}.
\label{four7} \eeq
The first term  in the right-hand side of Eq.~(\ref{four5})
can be interpreted as the adiabatic contribution to the spectrum, while 
the second one corresponds to entropic fluctuations. 
Until the end of the first stage of inflation
(when $t_1 \simeq t_{1f}$ according to Eq.~(\ref{three2})),
the functions $V_1(\sxa)$, $V_2(\sxb)$ are the potentials of the two
disconnected 
fields $\sxa$, $\sxb$ and can be derived from 
Eqs.~(\ref{two3}), (\ref{two7a}):
\beq
V_1(\sxa)=&~\mu_1^4+
\frac{M_1}{16 \pi^2}  \lx_1^2 \mu_1^4
\left[ \ln\left(
\frac{\lx_1^2 \sxa^2}{2 \Lambda_1^2} 
\right) 
+ \frac{3}{2}
\right],
\nonumber \\
V_2(\sxb)=&~\mu^4_2
+ \frac{M_2 }{16 \pi^2}  
\left( \lx_2 \mu_2^2 + g \mu_1^2\right)^2
\left[ \ln\left(
\frac{\lx^2_2 \sxb^2}{2 \Lambda_2^2} 
\right) 
+ \frac{3}{2}
\right].
\label{four78} \eeq
However, the subsequent evolution
of the entropic contribution is very complicated, due to the
growth of the $\pha$ field and the possible decay of 
$\sxa$, $\pha$ into lighter species. For this reason, we concentrate
on the adiabatic contribution, which can be followed until
today. We point out, however, that for our choice of parameters of the model,
the total spectrum of adiabatic perturbations 
is dominated by the fluctuations of the $\sxb$ field, with 
the contribution of $\sxa$ being negligible (see below). Since the entropic 
contribution arises because of the simultaneous fluctuations of
at least two fields, the dominance of the $\sxb$ fluctuations
indicates that the entropic contribution to the spectrum might be
subleading.

Assuming that the field fluctuations are random gaussian variables, 
we obtain for the spectrum of adiabatic density perturbations 
\be
\delta_H^2(k)= \frac{1}{25\pi^2}
\left[ \frac{H^2}{\mpl^4}
\left\{ 
\left( \frac{V_1}{V_1'} \right)^2
+ \left( \frac{V_2}{V_2'} \right)^2
\right\}  \right]_{k=aH},
\label{four8} \ee
where primes on $V_1(\sxa)$, $V_2(\sxb)$
denote derivatives with respect to $\sxa$, $\sxb$ respectively.
We assume that 
the evolution of $\sxb$ is negligible during the first
stage of inflation (so that $\sxb \simeq \sx_{2i}$ during this
whole stage). This is guaranteed if the ratio of Eq.~(\ref{three5}) 
is much
smaller than 1.
We also concentrate on the case $V_1/V_1' \ll V_2/V_2'$, for which the
metric perturbations are mainly generated by the fluctuations of
the $\sxb$ field during both stages of inflation.
This requires
\be
\frac{\sx_{1i}}{M_1} \frac{1}{ \lx_1^2} \ll 
\frac{\sx_{2i}}{M_2} \left(
\frac{\mu_2^2}{\lx_2 \mu_2^2+g\mu^2_1} \right)^2
\label{four9} \ee
and leads to
\be
\delta_H(k) \simeq
\frac{8\pi}{5M_2}
\left( \frac{\mu_2^2}{\lx_2 \mu_2^2+g \mu_1^2} \right)^2
\frac{H_1\sx_{2i}}{\mpl^2}.
\label{four10} \ee
The predicted spectrum is scale invariant with a spectral index
$n=1$ to a very good accuracy. It provides a good approximation 
to the primordial spectrum for $k < k_2$. It should be contrasted with
the spectrum predicted by the model of Refs.~\cite{pol1,pol2,pol3},
which has a logarithmic dependence $\propto \ln^{1/2}(K/k)$ for 
small $k$.

During the second stage of inflation the metric perturbations are
generated by the $\sxb$ field and the spectrum is
\be
\delta_H(k) \simeq
\frac{8\pi}{5M_2 \lx_2^2}
\frac{H_2\left[
\sxb
 \right]_{k=aH_2}}{\mpl^2}.
\label{four11} \ee
The field $\sxb$ evolves from $\sx_{2i}$ to $\sx_{2f}$
during this stage and the resulting spectrum 
has a spectral index different from 1. 
The scale dependence is given by the relation
\be
\left[ \frac{\sxb}{\mpl} \right]_{k=aH_2}^2
=~
\frac{\sx_{2i}^2}{\mpl^2} +
\frac{M_2 \lx^2_2}{4 \pi^2}
\left[
N_{1tot}+N_{int}-\ln \left( \frac{k}{H_0} \right)-
\ln \left( \frac{H_1}{H_2}\right)
\right].
\label{four12} \ee
The resulting spectral index is
\be
n-1 = \frac{d\ln\delta^2_H}{d\ln k}
=-\frac{M_2 \lx_2^2}{4 \pi^2}  \frac{\mpl^2}{\sxb^2(k)},
\label{four13} \ee
with $\sxb(k)$ given by Eq.~(\ref{four12}).
We are interested in scales that crossed outside the horizon during the
beginning of the second stage of inflation. For them 
the spectral index is given by 
the above expression with $\sxb(k) \simeq \sx_{2i}$.

The calculation of the form of the spectrum for scales $k_2<k<k_1$ that
re-enter the horizon during the intermediate stage is prohibited by the
extremely complicated nature of this stage.
This is due to the presence of
two massive oscillating and decaying fields $(\sxa,\pha)$.
However, we do not expect any strong features (such as spikes
\cite{spike})
in the spectrum associated with the intermediate stage. The situation is
very similar to the ``double inflation with a break'' of Ref.~\cite{pol1}.
In that case there is only one massive oscillating field with no decay 
channels, whose energy is dissipated through expansion. The detailed
calculation of the spectrum has not revealed any strong features.
In our model, no ``massless'' fields exist during the inflationary stages,
other than $\sxa$, $\sxb$ whose contribution to the spectrum we have
computed. The first stage of inflation ends for 
$\sxa \sim \sx_{1f} \gg \sx_{1ins}$, long before the orthogonal field
$\pha$ becomes massless or develops an instability; and
similarly for the second stage. 
A smooth interpolation between the two parts of the spectrum
that we have computed is likely, such as the one indicated by the data.

We also mention that the gravitational wave contribution to the spectrum
is not expected to be large. For our choice of the parameters of
the model (see next section), 
it is more than two orders of magnitude smaller than the scalar contribution
during the second stage of inflation. Due to the similarity of the
potentials of the first and second stage, we expect that the same is
true for the first stage of inflation as well.

Finally, we consider the effect of the subsequent inflationary stages
on the spectrum. The presence of a 
third stage, generated by an additional term  $S_3 
\left( -\mu_3^2 + \lx_3 \bar{\Phi}_3 \Phi_3 \right)$ in the superpotential,
could affect our previous discussion. In order to be more specific,
let us concentrate on the second and third inflationary stages. 
In the presence of additional superfields $\Phi_3$, $\bar{\Phi}_3$,  
the superpotential associated with the second inflationary stage 
has the more general form $S_2
\left( -\mu_2^2 + \lx_2 \bar{\Phi}_2 \Phi_2 + g' \bar{\Phi}_3 \Phi_3 \right)$.
For sufficiently small coupling $g'$, the potential that generates 
a slope along the flat direction during the second stage has a form analogous
to eq. (\ref{two7a}), i.e.
\be
\Delta V(\sxb,\sxc) \simeq~ 
\frac{M_2}{16 \pi^2}  \lx_2^2 \mu_2^4
\left[ \ln\left(
\frac{\lx_2^2 \sxb^2}{2 \Lambda_2^2} 
\right) 
+ \frac{3}{2}
\right]
+ \frac{M_3 }{16 \pi^2}  
\left( \lx_3 \mu_3^2 + g' \mu_2^2\right)^2
\left[ \ln\left(
\frac{\lx^2_3 \sxc^2}{2 \Lambda_3^2} 
\right) 
+ \frac{3}{2}
\right].
\label{two7aa} \ee
For simplicity we assume that
the value of $\sx_3$ does not change substantially during the
second stage. This is expected to be the case for
$\mu_2$ sufficiently larger than $\mu_3$.
The contribution of the fluctuations of the
$\sx_3$ field to the spectrum generated during the second stage 
can be estimated through the analogous of eq. (\ref{four8}). These 
fluctuations do not affect the spectrum if 
\be 
f^2/c^4 \ll 1,
\label{constr} \ee 
where
\beq
f = &
\frac{\sx_{3i}}{M_3 \lx_3^2} 
\frac{M_2 \lx_2^2}{\sx_{2i}},   
\nonumber \\
c = & 1+
\frac{g' \mu_2^2}{\lx_3~\mu_3^2}.
\label{constrpar} \eeq
The above constraint 
has implications for the total number of e-foldings $N_{3tot}$, given by 
Eq.~(\ref{three99}), that can
be generated during the third stage. 
For $\sx_{2f} \ll \sx_{2i}$, 
$\sx_{3f}\ll \sx_{3i}$, we can express $N_{3tot}$ as 
\be
N_{3tot} = N_{2tot} f~ \frac{\sx_{3i}}{\sx_{2i}}.
\label{constrr}\ee 
Large values of $N_{3tot}$ require values of $f$ that may be in conflict
with the constraint (\ref{constr}).
For  $g'=0$,
$N_{2tot}\sim 2-3$, $\sx_{2i}/\mpl=0.025$
(the value of this parameter that we use in the next section),
and allowing for
$f^2 \sim \frac{1}{4}$--$\frac{1}{3}$, 
a value of $\sx_{3i}$ slightly below 
$\mpl$ 
can lead to the required remaining $\sim 50$ e-foldings. A non-zero 
value of $g'$ leads to $c>1$ and permits even larger values of $f$
and, therefore, smaller values of $\sx_{3i}$.

The discussion of the first stage can be carried out in an analogous way, 
even though the constraints are more complicated. We expect that, 
similarly to above, 
there are regions of parameter space for which the primordial spectrum 
of the first stage is not affected by the presence of the additional fields. 
Moreover, it is conceivable that several similar inflationary stages generate 
a total number of e-foldings $\sim 50$, in which case the resulting bounds
are less stringent. As we have 
discussed in the introduction
and section 2, we have in mind a picture of multiple bursts of inflation, 
driven by different sectors of the theory, 
with a sequence of scales starting near $\mpl$ and continuing down to the scale
implied by COBE, or even to lower energy scales. 
In this work we present only a simplified picture of two of these stages
that have direct observational consequences. 

Another possibility is that the later stages of inflation are associated
with fields that are massive during the first stages or located near the
minima of their potential, and, therefore,
do not affect the spectrum. These fields are
displaced from their initial position during the reheating and trigger
inflation only when the temperature becomes sufficiently low. 
For such models the constraints discussed above 
do not apply. The scenario of
thermal inflation \cite{thermal} is a typical example.

The observational consequences of the third inflationary stage
are difficult to estimate. One could try to compute the linear mass fluctuation
$\sigma(R)$ for $R \sim 1~h^{-1}{\rm Mpc}$ and compare with 
the abundance of galaxies at large redshifts \cite{R1}. However, the spectrum
for the relevant
region of $k$ is strongly affected by the intermediate stage between the
second and third inflationary stages, for which a computation is impossible.
The fact that the scale $\mu_3$ is largely 
unconstrained by previous considerations
makes it probable that an agreement with observations is feasible.

\setcounter{equation}{0}
\renewcommand{\theequation}{{\bf 6.}\arabic{equation}}

\section{Comparison with observations}

In this section we compare the predictions of our model with the 
experimental data from COBE and the APM galaxy survey. 
We do not consider the gravitational wave contribution to the spectrum,
as we expect it to be small.
We also concentrate on the adiabatic scalar contribution given by 
Eqs.~(\ref{four10})--(\ref{four13}).
We do not consider any early ionization scenarios or HDM.
We are interested in examining whether the COBE-DMR measurements 
can be made compatible with the data of galaxy surveys 
employing the smallest number of cosmological 
parameters.  
More specifically, we consider two cosmological scenaria:
1)
A CDM model, with zero cosmological constant, present Hubble parameter $h=0.5
~~(H_0 \equiv h~100~{\rm km/s/Mpc})$ and flat geometry $\Omega_{matter}=1$. 
We assume that the fraction of the critical density in
baryons is $\Omega_{b}=0.05$ and in cold dark matter $\Omega_{CDM}=0.95$. 
The helium abundance is $Y_{He}=0.24$, while we do not consider any  massive 
neutrinos. 2)
The same model with $\Omega_{\Lx}=0.5$, $h=0.5$,
$\Omega_{b}=0.05$ and $\Omega_{CDM}=0.45$. 

In both the above scenaria, 
we use the following parameters for the model of Section 3:
$\lx_1=1$, $\lx_2=0.1$, $g=7.1 \times 10^{-3}$, 
$\mu_1/\mpl=3.9 \times 10^{-3}$,
$\mu_2/\mpl=8.8 \times 10^{-4}$,
$\sx_{1i}/\mpl=0.36$, $\sx_{2i}/\mpl=0.025$.
For the purpose of this section, 
the parametes $M_{1,2}$, which determine
the dimensionalities of the representations
of the groups ${\rm \cal M}_{1,2}$ to which the superfields
$\Phi_{1,2}$ belong, can be absorbed in a redefinition
of $\lx_1$, $\lx_2$, $g$. For this reason we set $M_1=M_2=1$.
For the above choice of parameters the 
observable part of the first stage of inflation 
generates $N_{1tot}\simeq 5$ e-foldings and a spectrum with
a spectral index $n \simeq 1$. For the second stage of inflation,
Eq.~(\ref{three9}) predicts $N_{2tot}\simeq 2$ e-foldings.
However, a more accurate 
numerical integration of the field evolution equations
gives $N_{2tot}\simeq 3$.
The spectral index relevant for the scales that cross
outside the horizon in the beginning of the second
stage of inflation is given by Eq.~(\ref{four13}) with
$\sxb(k) \simeq \sx_{2i}$ and 
is $n \simeq 0.6$. The intermediate stage affects the
scales $k_2<k<k_1$ with 
$k_1/k_2 \simeq 4.4$ according to Eq.~(\ref{four4}).
The values of the power spectrum for 
$k_2$ and $k_1$ are determined by Eqs.~(\ref{four10}),
(\ref{four11}) and satisfy
$P(k_2)/P(k_1)\simeq 11.8$.

The initial spectrum of fluctuations arising in our model 
generates matter and 
radiation perturbations, which, after amplification,  
give rise to the observed large scale structure 
and the anisotropies of the cosmic microwave background.  
In the following,  
we compare the theoretical predictions of our model 
with the measured anisotropies
of the cosmic microwave background and the observed  
clustering in the galaxy distribution. 
For the comparison with the CMB data, we assume a form of
the power spectrum in the range $k_2<k<k_1$ that interpolates
smoothly between the spectrum in the ranges 
$k>k_1$ and $k<k_2$.

\subsection{Angular power spectrum of CMB anisotropies}

The cosmic microwave background (CMB) radiation, last scattered at the epoch of
decoupling, follows a 
blackbody distribution to high accuracy
\cite{bb}, with a temperature almost independent of direction, 
$T=2.728\pm 0.002 K$.
As it was measured by the DMR experiment on the COBE satellite, the  CMB
radiation has a tiny variation in intensity at fixed frequency, equivalently
expressed as a variation $\Delta T$ in the temperature: 
$\Delta T/T < 10^{-5}$. The
4-year COBE data is fitted by a scale-free spectrum, with spectral index
$n=1.2\pm .3$ 
and a quadrupole normalization
$Q_{rms}=15.3^{+3.7}_{-2.8} \mu K$ \cite{cobe}.

In Fig.~2 we compare our theoretical predictions for the angular 
power spectrum of 
CMB anisotropies against the most recent CMB flat-band power measurements
(we have used the data of Table 1 in Ref.~\cite{lb}). 
The coefficients $C_\ell$ correspond to the expansion
of the angular correlation function in powers of the Legendre 
polynomials $P_\ell$.
For large $\ell$, the angular size (in radians) 
of a feature in the sky is
related to the order of the multipoles that dominate it
by $\ell\sim 1/\theta$.
The angular power spectrum of CMB anisotropies is normalized
to the four-year data from COBE-DMR \cite{norm}. 
In order to 
calculate the predictions of our model for 
$C_\ell$ we have used the code 
CMBFAST of Seljak and Zaldarriaga \cite{cmbfast}.

In our model, the first stage of inflation leads to a scale-invariant
Harrison-Zel'dovich \cite{hz} (spectral index $n= 1$) spectrum of 
perturbations.
The intermediate stage and the second stage of inflation 
affect only the highest multipoles in Fig.~2. 
For this reason, our
prediction for the first acoustic peak 
in the case $\Omega_\Lx=0$ (short-dashed line)
is similar to that of the standard CDM (SCDM) model (solid line)
with a Harrison-Zel'dovich spectrum at all scales. 
Only for multipoles $l \sim 500$--1000 a drop of the values of 
$C_l$ is observed relative to the standard CDM model.
This should be contrasted with the 
double-inflationary model of Refs.~\cite{pol1,pol2,pol3},
which predicts a
first acoustic peak that is too low with respect to the Sachs-Wolfe plateau.
This problem can be traced to the 
logarithmic dependence $\propto \ln^{1/2}(K/k)$ of the spectrum for 
small $k$ in this model. Better agreement with the data is obtained
for our model with $\Omega_\Lx=0.5$ (long-dashed line).

\subsection{Angular catalogues of galaxy positions}

Angular catalogues of galaxy positions provide strong constraints
on theories of structure formation in the universe. In this
subsection we compare the
predictions of our model with the power spectrum 
of galaxy clustering derived from the angular APM galaxy survey 
\cite{apm}--\cite{brdat2}. 
Redshift surveys are more noisy than the APM survey at large scales
and subject
to more uncertainties due to the distortion of the pattern of galaxy
clustering by the peculiar motions of galaxies.

As we have discussed in Section 2, 
the effects of the non-linear evolution
of matter fluctuations (when the density contrast becomes of order 1) 
must be removed before a comparison is possible with
the linear spectrum. This is usually done through $N$-body simulations
of clustering. 
Baugh and Gazta\~naga \cite{brdat2} have estimated the 
linear spectrum that gives rise to the APM spectrum for
a spatially flat universe with critical density and 
zero cosmological constant. Their estimate
is well fitted by the curve of Eq.~(\ref{lfit})
This result is in agreement with the
spectrum obtained through the application of the formula of
Jain {\it et al.} \cite{jmw} for the relation between the linear and
non-linear spectra. The 
formula of Peacock and Dodds \cite{pd} does not lead to 
very good agreement with the results of the
$N$-body simulations \cite{linear}. 
The effects of the non-linear evolution are significant for
$k \gta 0.2 h {\rm Mpc}^{-1}$.

For the comparison of our predictions with the data, we 
calculate the transfer function $T(k)$ to better accuracy than
the one provided by Eq.~(\ref{tfit}), by employing the code CMBFAST of 
Seljak and Zaldarriaga \cite{cmbfast}.
The linear power spectrum 
is related to the
density perturbation at horizon crossing through \cite{report}
\be
\De^2(k)=\frac{k^3 P(k)}{2 \pi^2} 
= \de_H^2(k)~\left ( {k \over H_0}\right)^4 ~T^2(k)~,
\ee
where the dimensionless quantity $\De^2(k)$ stands for the 
mass variance per unit
interval in $\ln k$ (i.e., $\De^2(k) \equiv d\si^2_{mass}/d\ln k$).

In Fig.~3 we display the power spectrum $P(k)$ in units of 
$({\rm Mpc}/h)^3$ as a function of  $k$ in
units of $h/{\rm Mpc}$ for our model with $\Omega_\Lx=0$ and 
$k_2= 0.06~ h {\rm Mpc}^{-1}$.
For scales $k <  k_2$ the
spectral index has been taken $n=1$, while for scales 
$k> k_1$ the spectrum 
is tilted with spectral index $n = 0.6$. For scales
$k_2 <k< k_1$ the complicated nature of the intermediate stage makes 
the calculation of the power spectrum very difficult.
The prediction of our model for the linear spectrum is given 
by the short-dashed line.
The points with error bars correspond to the data from the
APM angular galaxy catalogue \cite{data1,brdat2}. 
The fit of Eq.~(\ref{lfit}) for the
linear spectrum derived from these data is given by the solid line.

We observe good agreement of our predictions with the linear spectrum
expected from the APM data. This agreement should persist
for the scales $k_2 <k< k_1$, as no strong new feature in the spectrum
is expected in this interval (see discussion at the end of Section 5). 
We point out that we have not assumed 
any biasing in the galaxy distribution with respect to the underlying
mass fluctuations (bias parameter $b_g=1$). Allowing for other values
of $b_g$ provides an additional degree of freedom that can improve
the agreement of our predictions with observations.

The power spectrum for our model with $\Omega_\Lx=0.5$ is 
given by the long-dashed line in Fig.~4. We compare again with
the estimate of Eq.~(\ref{lfit})
for the linear spectrum (solid line).
The reason is that for a flat geometry of the universe we
do not expect a significant modification of this estimate.
The formula of Peacock and Dodds \cite{pd}
for the relation between the linear and non-linear spectra 
predicts only a small difference
between the spectra for $\Omega_{matter}=1$, $\Omega_\Lx=0$ and
for $\Omega_{matter}=0.5$, $\Omega_\Lx=0.5$.
We observe a good agreement between our prediction and the linear spectrum
in the range 
$k \geq 0.26 ~h{\rm Mpc}^{-1}$. On the other hand,
our prediction is significantly above the linear
spectrum for $k \leq 0.06 ~h{\rm Mpc}^{-1}$.
However, the observational data in this last range 
are dominated by large errors. This makes the estimate for the linear spectrum
less accurate.

For values of $\Omega_\Lx$ above 0.5, the discrepancy between our prediction
and the APM spectrum at large scales is enhanced. Reconciling the prediction 
with the data can be achieved either through
the consideration of a bias parameter different from 1 for the 
galaxy distribution or a spectral index smaller than 1 for the first stage 
of inflation.

As a final test of our model, we turn to the calculation of the
parameter $\sei$, which is a common measure of large-scale clustering.
It is defined as the variance of the density field smoothed over
a radius $R=8~h^{-1}{\rm Mpc}$ 
\be
\sx^2_8 = \frac{1}{H_0^4}
\int_0^{\infty} W^2(k R) \delta_H^2(k) T^2(k) k^3 dk.
\label{s8} \ee
The ``top-hat'' smoothing function $W(kR)$ is given by 
\be
W(kR) = 3 \left[\frac{\sin(kR)}{(kR)^3}-\frac{\cos(kR)}{(kR)^2} \right].
\label{smooth} \ee
For $n\sim 0.6$ the largest contribution to the integration of Eq.~(\ref{s8})
comes from $k \sim 1/R = 0.125~h {\rm Mpc}^{-1}$. This corresponds to the
tilted part of the spectrum near $k_1$, for which we have used $n=0.6$ 
in the calculation of the
results presented in Figs.~3 and 4. However, as Fig.~1 
indicates, the effective spectral index may be even smaller for
$k$ slightly below $k_1$. For this reason we have estimated $\sei$ allowing for
a local variation of $n$ in the interval [0.2, 0.8]. We find that 
$\sei$ is not very sensitive to the value of $n$. Instead, it is
determined by the normalization of the spectrum at $k\simeq k_1$,
which is fixed by 
$P(k_2)/P(k_1)\simeq 11.8$ for our choice of parameters of the model.
The predicted values of
$\sei$ are near 0.75, both for $\Omega_\Lx=0$ and $\Omega_\Lx=0.5$.
They are in good agreement with
the values deduced from 
the abundances of rich clusters of galaxies:
$\sigma_8\simeq 0.6 \pm 0.2$ for $\Omega_{matter}=1$, $\Omega_\Lx=0$
\cite{wef93}, and $\sigma_8\simeq 0.85 \pm 0.3$ 
for $\Omega_{matter}=0.5$, $\Omega_\Lx=0.5$ \cite{viana}. 

\setcounter{equation}{0}
\renewcommand{\theequation}{{\bf 7.}\arabic{equation}}

\section{Summary and conclusions}

In this paper we considered the possibility of a multiple-stage
inflationary scenario within the context of supersymmetric 
hybrid inflation. This framework has the nice feature that 
the observable part of inflation takes place for field values
below the Planck scale. This permits the discussion of 
corrections to the tree-level picture (such as 
supergravity corrections) in terms of 
controlled expansions in powers of fields in units of $\mpl$. 
Inflationary models such as the one we considered have
been discussed in Ref.~\cite{mecostas}, and have been shown
to survive the supergravity corrections in Ref.~\cite{megeorge}. 

One motivation for the consideration of a scenario with more than
one stages of inflation arises from the discrepancy between
the inflationary energy scale implied by the COBE data and the 
Planck scale at which classical general relativity becomes applicable.
This discrepancy results in the necessity of an extremely fine-tuned
field configuration at the end of the Planck era for the onset of
hybrid inflation \cite{nikos}. The problem can be avoided if
inflation takes place in two stages \cite{mecostas}.
The first stage has a typical scale $\sim \mpl$ and 
occurs naturally. The second has a characteristic
scale much below $\mpl$ and generates the density perturbations 
that result in the cosmic microwave background
anisotropy measured by the DMR experiment on the COBE satellite. 
It is the logical next step to contemplate the 
possibility of a multiple-stage inflationary scenario, with a
sequence of scales between $\mpl$ and close or below 
the one implied by COBE.

Another motivation for the consideration of a multiple-stage scenario
stems from the need to reconcile the power spectrum of density
perturbations at large scales 
(obtained through the COBE-DMR measurements for the anisotropies of the 
cosmic microwave background) with the observed power spectrum at small 
scales (deduced from galaxy surveys). The standard CDM model
with a flat (spectral index $n=1$) initial spectrum of adiabatic
perturbations predicts too much power at small scales, 
if one normalizes it to the
COBE data at large scales. An interesting possibility in order to
resolve this discrepancy is to consider power spectra with a break.  
A strong indication for the location of the break is provided by the 
form of the power spectrum 
of galaxy clustering derived from the APM galaxy survey
\cite{data1,peac,subir,brdat2}.
At $k_b \simeq 0.05~h~{\rm Mpc}^{-1}$ the data support a sharp drop of the
spectral index of the primordial spectrum, from a value $n\simeq 1$ to 
an average value $n\sim 0.6$. This break has been linked to a possible phase
transition between two of the inflationary stages in a
multiple inflationary scenario \cite{subir}.

In this paper we presented a model of supersymmetric hybrid inflation
that generates a spectrum with a break. As we argued in 
Section 2, the large deviation of the spectral index from 
1 for $k \lta k_b \simeq 0.05~h~{\rm Mpc}^{-1}$
results in a small number of e-foldings generated by the stage of
inflation relevant for this range of scales. 
A picture of multiple short bursts of inflation, with significant 
variations of the spectral index of the primordial spectrum, seems 
probable. 

We computed the spectrum of adiabatic density perturbations
for the range of scales 
that are relevant for the COBE data and the APM galaxy survey.
This requires the explicit discussion of the first two observable
stages of inflation, which generate $\sim 8$ e-foldings. 
The total number of e-foldings must be $\sim 60$ for the flatness and 
horizon problems of
standard cosmology to be resolved. However, the stages 
responsible for the last $\sim 50$ e-foldings generate density perturbations
at scales that 
are strongly affected by the non-linear evolution. As a result, it is very
difficult to extract information on their characteristics.
 
In our model, the adiabatic density fluctuations 
during both stages of inflation originate in the quantum fluctuations
of the same field. The first stage of inflation generates 
$\simeq 5$ e-foldings and a spectrum with index $n = 1$
for scales $k \lta 0.06~h~{\rm Mpc}^{-1}$.
The second stage generates 
$\simeq 3$ e-foldings and a spectrum with index $n \simeq 0.6$ 
for scales $k \gta 0.26~h~{\rm Mpc}^{-1}$.
The calculation of the spectrum for 
$0.06~h~{\rm Mpc}^{-1} \lta k \lta 0.26~h~{\rm Mpc}^{-1}$
is prohibited by a very complicated intermediate stage of normal expansion.
However, it probable that 
no significant new feature arises in this range of
scales, as we discussed at the end of Section 5.
 
We considered two scenaria:\\
a) A CDM model, with zero cosmological constant, present Hubble 
parameter $h=0.5 (H_0 \equiv h~100~{\rm km/s/Mpc})$ and flat geometry -- 
$\Omega_{matter}=1$. We assumed that the fraction of the critical density in
baryons is $\Omega_{b}=0.05$ and in cold dark matter $\Omega_{CDM}=0.95$.\\
b) A scenario with 
$\Omega_{\Lx}=0.5$, $\Omega_{b}=0.05$ and $\Omega_{CDM}=0.45$.\\
The comparison of our predictions with observations is given in Figs.~2--4. 
For the first scenario, 
the angular power spectrum of CMB anisotropies (Fig.~2) is very 
similar to the one with 
a Harrison-Zel'dovich spectrum at all scales.
Only for multipoles $l \sim 500$--1000 a drop of the values of 
$C_l$ is observed relative to the standard CDM model. This should be contrasted
with the predictions of a model of double chaotic inflation, for which
the first Doppler peak is significantly suppressed \cite{pol3}. 
The second scenario leads to a better agreement of the generated spectrum
with the data.
The power spectrum predicted by the first scenario compares well with the
linear power spectrum deduced from the APM data (Fig.~3).
The agreement is less satisfactory for the second scenario (Fig.~4), even
though the discrepancy appears in the region of large data errors.
For both scenaria, our predictions for $\sei$ are in agreement with the
values deduced from the abundances of rich clusters of galaxies.

As a closing remark, we point out that the questions of the 
presence and the nature of a feature in the power spectrum of the 
galaxy distribution are not settled. Several interpretations and
explanations of the data exist (e.g., see Refs.~\cite{pol4,barand}). 
Our work provides a possible explanation of the observed feature
within a multiple-stage inflationary 
scenario in the context of supersymmetric hybrid 
inflation. 

\vspace{0.5cm}
\noindent
{\bf Acknowledgements}: 
It is a pleasure to thank J. Adams, N. Deruelle, E. Gazta\~naga 
J. Martin, A. Melchiorri, D. Polarski and S. Sarkar for useful discussions, 
and U. Seljak for help with CMBFAST. M.S. would also like to thank 
C.M. Viallet.
One of us (M.S.) thanks the D\'epartement d'Astrophysique 
Relativiste et de Cosmologie, Observatory of Paris (Meudon), for hospitality 
during the preparation of this work. 
The work of N.T. was supported by the E.C. under TMR contract 
No. ERBFMRX--CT96--0090.

\newpage

\begin{figure}[htb]
\vspace{1.cm}
\centerline{\psfig{figure=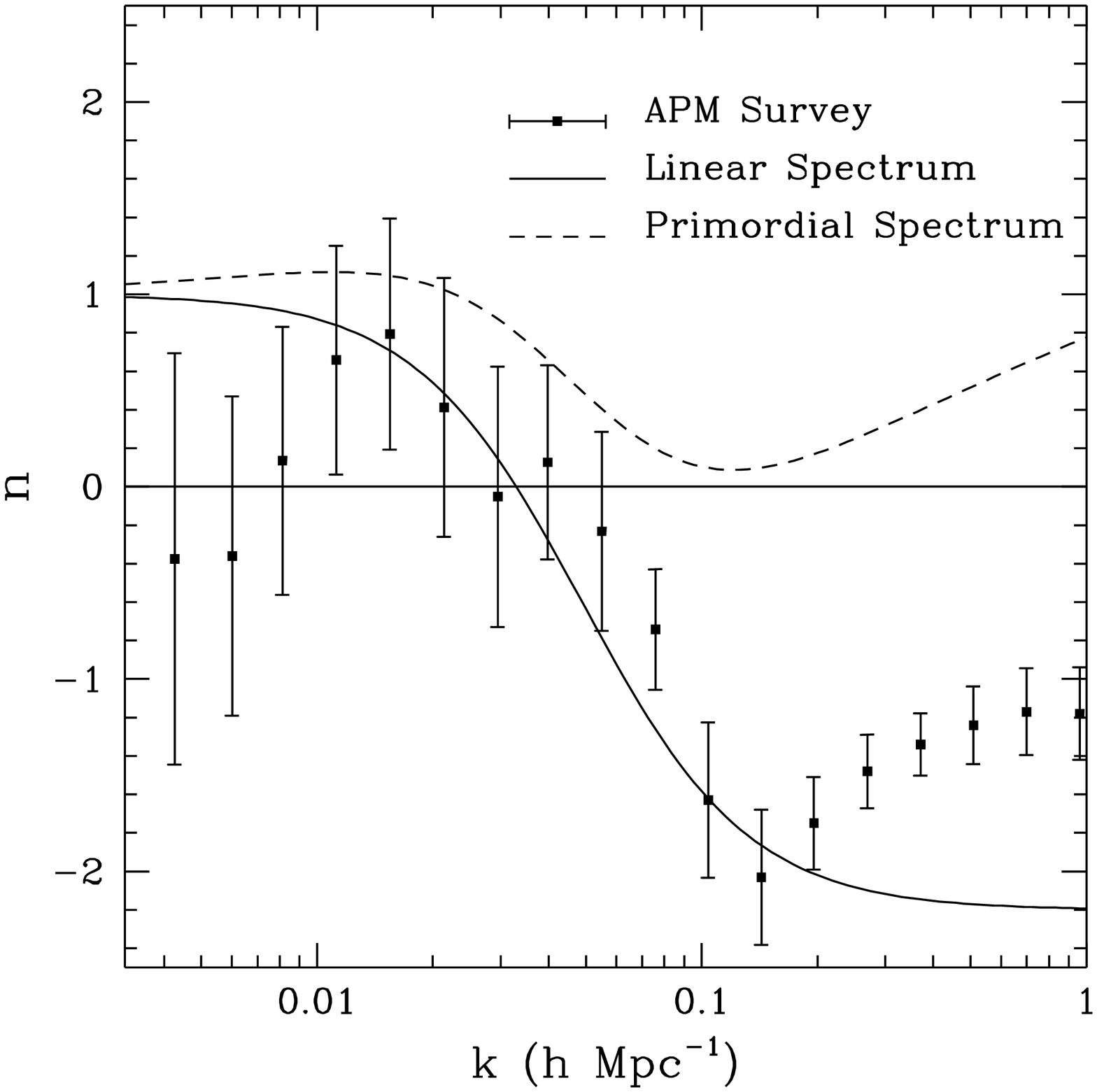,width=12cm}}
\caption{
The logarithmic slope of:
(a) The power spectrum
recovered from the APM angular galaxy catalogue (points with error bars).
(b) The estimated 
linear spectrum that gives rise to the observed spectrum (solid line).
(c) The estimated primordial spectrum (dashed line).
(After Refs.~\cite{subir,brdat2,linear}).}
\end{figure}

\newpage

\begin{figure}[htb]
\vspace{1.cm}
\centerline{\psfig{figure=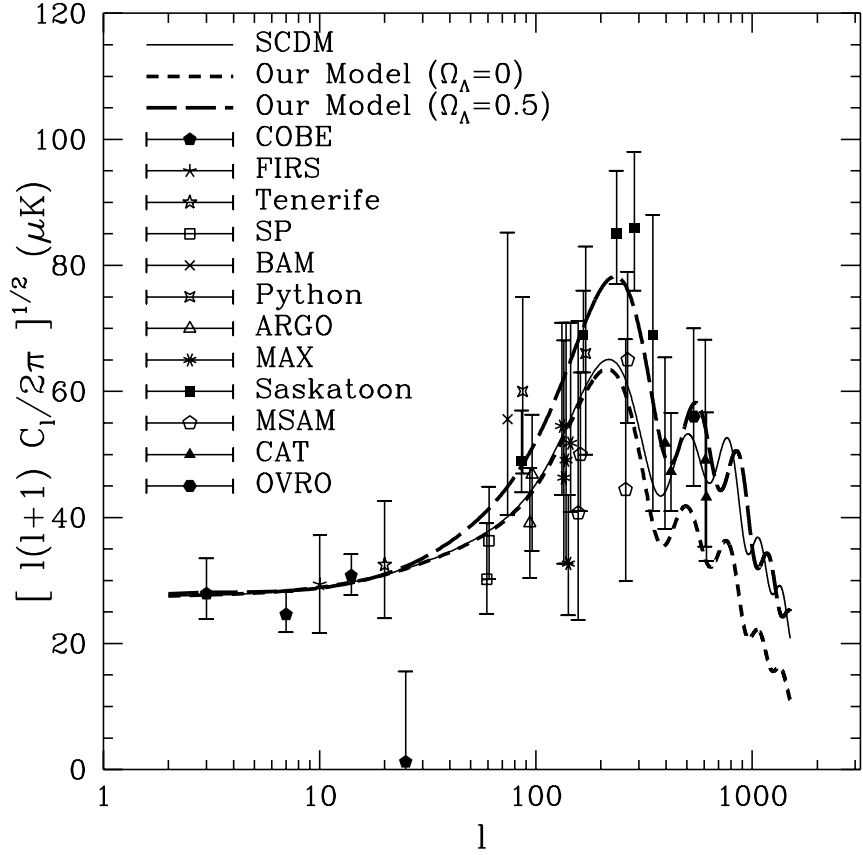,width=12cm}}
\caption{Theoretical predictions for the angular power spectrum of CMB 
anisotropies in our multiple-stage inflation model, against the
most recent CMB flat-band power measurements. 
We plot
$\sqrt{\ell (\ell +1) C_{\ell}/(2\pi)}$ in units of $\mu K$ 
versus the multipole
moment $\ell$. The solid line corresponds to the standard CDM model 
($\Omega_\Lx=0$, $n=1$). The short-dashed and long-dashed lines 
correspond to the 
spectra in our model for $\Omega_\Lx=0$ and $\Omega_\Lx=0.5$
respectively. 
}
\end{figure}

\newpage

\begin{figure}[htb]
\vspace{1.cm}
\centerline{\psfig{figure=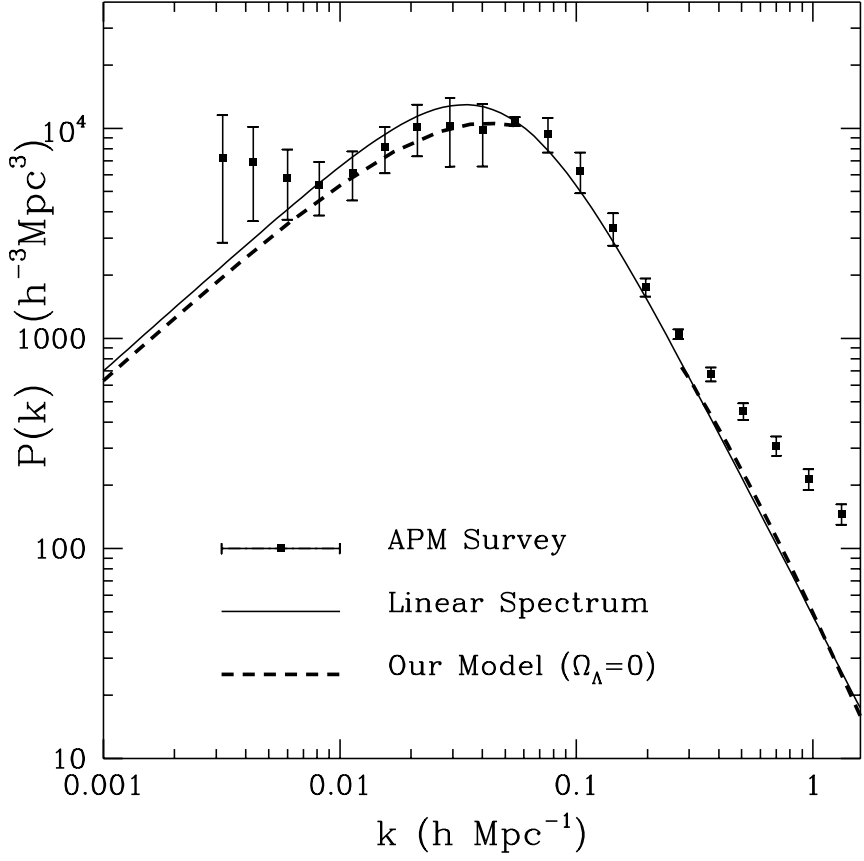,width=12cm}}
\caption{Power spectrum generated in our model
(short-dashed line) for $\Omega_\Lx=0$, plotted together with the 
APM power spectrum (data points)
and the linear spectrum of Eq.~(\ref{lfit})
(solid line).
The spectrum is flat for $k \leq 0.06 ~h{\rm Mpc}^{-1}$ and
tilted with $n=0.6$ for $k \geq 0.26 ~h{\rm Mpc}^{-1}$.}
\end{figure}

\newpage

\begin{figure}[htb]
\vspace{1.cm}
\centerline{\psfig{figure=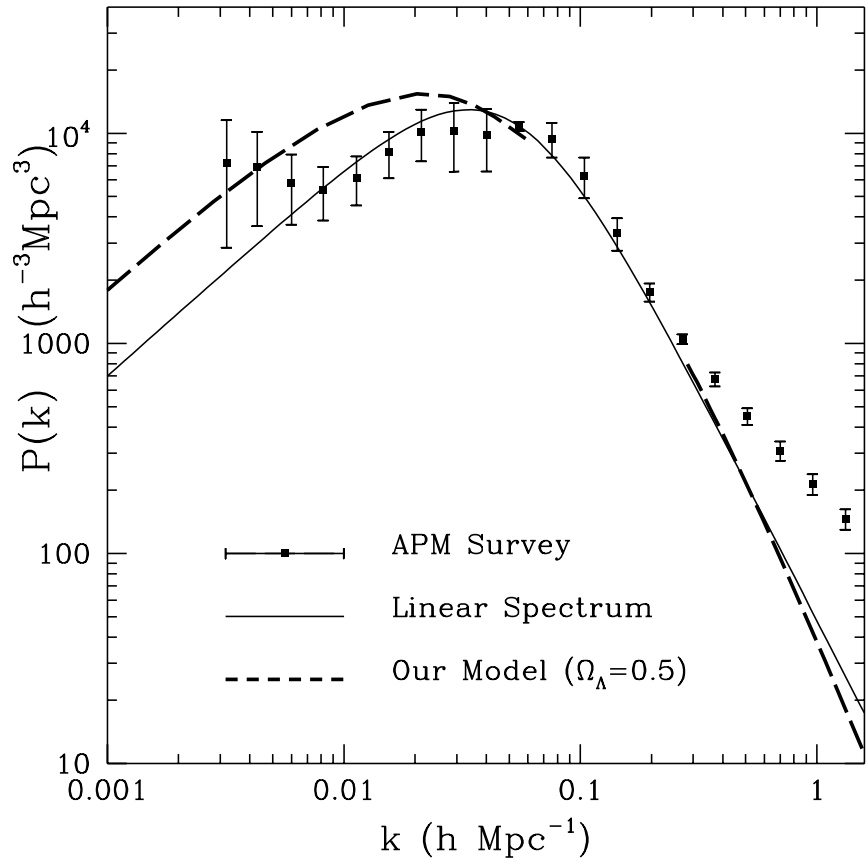,width=12cm}}
\caption{Same as in Fig.~3 for $\Omega_\Lx=0.5$.}
\end{figure}

\end{document}